\documentclass[journal,final]{IEEEtran}

\usepackage{xcolor}
\usepackage{graphicx}
\usepackage{tikz}
\usepackage{color}
\usepackage{amsmath}
\usepackage{subfig}
\usepackage{algorithm}
\usepackage{algorithmic}
\usepackage{amssymb}
\graphicspath{{./Fig/}}
\usepackage{math}
\usepackage{caption}
\usepackage{array}
\usepackage{enumitem}
\usepackage{balance}

\usepackage[]{review}
\setcoverletter{coverletter-R1.tex}
\setrevision{1}

\begin{document}


\title{Multichannel Online Dereverberation Based on Spectral Magnitude Inverse Filtering
}
\author{Xiaofei Li, Laurent Girin, Sharon Gannot and Radu Horaud
\thanks{X. Li and R. Horaud are with INRIA Grenoble Rh\^one-Alpes, Montbonnot Saint-Martin, France. }
\thanks{L. Girin is with GIPSA-lab and with Univ. Grenoble Alpes, Saint-Martin d'H\`eres, France.}
\thanks{Sharon Gannot is with Bar Ilan University, Faculty of Engineering, Israel. }
\thanks{This work was supported by the ERC Advanced Grant VHIA \#340113.}
}


\maketitle

\begin{abstract}
This paper addresses the problem of multichannel online dereverberation. 
The proposed method is carried out in the short-time Fourier transform (STFT) domain, and for each frequency band independently. In the STFT domain, the time-domain room impulse response is approximately represented by the convolutive transfer function (CTF). The multichannel CTFs are adaptively identified based on the cross-relation method, and using the recursive least square criterion. Instead of the complex-valued CTF convolution model, we use a nonnegative convolution model between the STFT magnitude of the source signal and the CTF magnitude, which is just a coarse approximation of the former model, but is shown to be more robust against the CTF perturbations. Based on this nonnegative model, we propose an online STFT magnitude inverse filtering method. The inverse filters of the CTF magnitude are formulated based on the   multiple-input/output inverse theorem (MINT), and adaptively estimated based on the gradient descent criterion. Finally, the inverse filtering is applied to the STFT magnitude of the microphone signals, obtaining an estimate of the STFT magnitude of the source signal. Experiments regarding both speech enhancement and automatic speech recognition are conducted, which demonstrate that the proposed method can effectively suppress reverberation, even for the difficult case of a moving speaker.  
\end{abstract}


\section{Introduction}
\label{sec:introduction}

This paper addresses the problem of multichannel online dereverberation of speech signals, emitted by either a static or a moving speaker. The objective of dereverberation is to improve speech quality/intelligibility for human listening or for automatic speech recognition (ASR). \addnote[overview]{1}{In the REVERB challenge \cite{kinoshita2016}, a number of dereverberation methods were benchmarked, which showed that both speech quality (naturalness, distortion, perceived reverberation, etc.) and ASR performance can be improved by dereverberation, and that larger the number of microphones better the improvement. As for ASR, \cite{li2017acoustic,caroselli2017,drude2018,heymann2018} show that, even for an advanced ASR back-end with multi-condition training to account for the reverberation effect, a standalone dereverberation front-end is still helpful. The influence of reverberation on speech intelligibility was analyzed in \cite{kressner2018,xia2018,santos2014,santos2014improved} for both normal- and hearing-impaired listeners. It was shown that, in office rooms, reverberation alone does not severely degrade speech intelligibility for normal-hearing listeners, while it does for hearing-impaired listeners. Under noisy conditions, reverberation significantly degrades speech intelligibility for both normal- and hearing-impaired listeners. It was shown in \cite{warzybok2014} that, for normal-hearing listeners,  dereverberation indeed improves the  tolerance of listeners to noise. Compared to normal-hearing listeners, \cite{zhao2018} showed that  speech intelligibility for hearing-impaired listeners can be prominently improved by dereverberation} .
The output of a dereverberation system may include some early reflections, since they deteriorate neither speech quality nor speech intelligibility \cite{arweiler2011}.

Multichannel dereverberation includes the following different techniques. Spectral enhancement techniques \cite{habets2009,schwarz2015,kuklasinski2016},  which are performed in the short-time Fourier transform (STFT) domain, remove  late reverberation by spectral subtraction. 
To iteratively estimate the room filters and the speech source signal, other techniques minimize a cost function between the microphone signal(s) and a generative model thereof (or equivalently maximize an objective function).
The generative model here mainly indicates the convolutive model between  the room filters and the source signal, and sometimes the source signal is assumed to be generated by a random process. These techniques are also usually applied in the STFT domain, where the time-domain RIR is represented by a subband convolutive transfer function (CTF). An expectation-maximization (EM) algorithm is used in \cite{schwartz2015} to maximize the likelihood of the microphone signals. The idea is extended to joint dereverberation and source separation in \cite{li2017waspaa}. In \cite{mirsamadi2014,mohammadiha2016,baby2017}, a nonnegative convolution approximation is assumed, namely the STFT magnitude of the microphone signal is approximated by the convolution between the STFT magnitude of  the source signal and the CTF magnitude. Based on this nonnegative model, tensor factorization \cite{mirsamadi2014},  iterative auxiliary functions \cite{mohammadiha2016} and iterative multiplicative update \cite{baby2017} are used to minimize the fit cost between the STFT magnitude of the microphone signal and its nonnegative generative model.  
Inverse filtering techniques aim at inverting the room convolution process and recovering the source signal. Depending on the way inverse filters are estimated, inverse filtering techniques can be classified into two groups:
\begin{itemize}[leftmargin=*]
 \item Linear prediction based techniques model the convolution with the RIR as an auto-regressive (AR) process. This AR process can be carried out either in the time domain or in the STFT domain. In the linear-predictive multi-input equalization (LIME) algorithm \cite{delcroix2007}, the speech source signal is estimated as the multichannel linear prediction residual, which however is excessively whitened. The whitening effect is then compensated by estimating the average speech characteristics. To avoid such whitening effect,  a prediction delay is used in the delayed linear prediction techniques \cite{kinoshita2009,nakatani2010}. These techniques only model late reverberation into the AR process and leave early reflections of the speech signal in the prediction residual. To account for the time-varying characteristics of speech,  the statistical model-based approach \cite{nakatani2010} iteratively estimates the time-varying speech variance and normalizes the linear prediction with this speech variance. This 
variance-normalized delayed linear prediciton method is also called weighted prediction error (WPE); 
\item Techniques based on system identification first blindly identify the room filters. Then, the corresponding inverse filters are estimated and applied on the microphone signals to recover the source signal. The cross-relation method \cite{xu1995} is a widely-used system identification method. Inverse filter estimation techniques include the multiple-input/output inverse theorem (MINT) method \cite{miyoshi1988} and some of its variants, such as channel shortening \cite{kallinger2006} and partial MINT \cite{kodrasi2013}. In \cite{li2016taslp,li2018taslp}, the cross-relation method was applied in the STFT domain for CTF estimation. Several variants of subband MINT were proposed based on filter banks \cite{weiss1999,gaubitch2009} or CTF model  \cite{li2018icassp,li2018taslp2}.
\end{itemize}

For dynamic scenarios with moving speakers or speech turns among speakers, an online dereverberation method is required. Based on the CTF model, an online likelihood maximization method was proposed in \cite{schwartz2015waspaa,schwartz2015online} using a Kalman filter and an EM algorithm. An online extension of LIME was proposed in \cite{yang2014} using several different adaptive estimation criteria, such as normalized least mean squares (LMS), steepest descent, conjugate gradient and recursive least square (RLS). \addnote[awpebest]{1}{RLS-based adaptive WPE (AWPE) \cite{caroselli2017,yoshioka2009,yoshioka2013,xiang2018} became a popular online dereverberation method. For example, it is used by the Google Home smart loudspeaker device \cite{li2017acoustic}}. In AWPE, the anechoic speech variance is estimated using a spectral subtraction method in \cite{yoshioka2013}, and is simply approximated by the microphone speech variance in \cite{yoshioka2009,caroselli2017,xiang2018}. In \cite{braun2016,braun2018}, a probabilistic model and a Kalman filter were used to implement the delayed linear prediction method, which can be seen as a generalization of the RLS-based AWPE. 
A class of adaptive cross-relation methods were proposed in \cite{huang2003} for online system identification, with the adaptive estimation criteria of normalized LMS and multichannel Newton method. Adaptive multichannel equalization methods were proposed in \cite{zhang2008,liu2011} based on time-domain MINT and gradient descent update. These methods reduce the computational complexity of the original MINT, however they were only used for offline multichannel equalization in static scenarios.

In our previous work \cite{li2018taslp}, a blind dereverberation method was proposed in batch mode for static scenarios. This method consists of a blind CTF identification algorithm and a sparse source recovery algorithm. The CTF identification algorithm is based on the cross-relation method. For source recovery, instead of the complex-valued CTF convolution model, we used its  nonnegative convolution approximation \cite{mirsamadi2014,mohammadiha2016,baby2017}, since the latter was shown to be less sensitive to the CTF perturbations than the former. More precisely, the STFT magnitude of the source signal is recovered by solving a basis pursuit  problem that minimizes the $\ell_1$-norm of the STFT magnitude of the source signal while constraining the fit cost, between the STFT magnitude of the microphone signals and the nonnegative convolution model, to be below a tolerance.

In the present work,  we propose an online dereverberation method. First, we extend the batch formulation of CTF identification in \cite{li2018taslp} to an adaptive method based on an RLS-like recursive update. The RLS-like method has a better convergence rate than the normalized LMS method used in \cite{huang2003}, which is crucial for its application in dynamic scenarios. 
\addnote[cpx1]{1}{This adaptive CTF identification is carried out in the complex domain, then the magnitude of the identified CTF is used for online inverse filtering, based on the nonnegative convolution model: the inverse filters of the CTF magnitudes are adaptively estimated and applied to the STFT magnitude of the microphone signals to obtain an estimate of the STFT magnitude of the source signal}.
The inverse filters estimation is based on the MINT theorem \cite{miyoshi1988}.  Due to the use of the nonnegative CTF convolution model, the proposed magnitude MINT is different from the conventional MINT methods, such as \cite{kallinger2006,kodrasi2013,li2018icassp}, mainly in aspect to that multichannel fusion and target response. Following the spirit of normalized LMS, we propose to adaptively update the inverse filters based on a gradient descent method.
In summary, the proposed method consists of two novelties i) an online RLS-like CTF identification technique, and ii) an online STFT-magnitude inverse filtering  technique. To the best of our knowledge this is the first time such procedures are proposed for online speech dereverberation.
Experimental comparison with AWPE shows that the proposed method performs better for the moving speaker case, mainly due to the use of the less sensitive magnitude convolution model. 

The remainder of this paper is organized as follows. The adaptive CTF identification is presented in Section~\ref{sec:ci}.
The online STFT magnitude inverse filtering method is presented in Section~\ref{sec:dereverberation}.  
Experiments with two datasets are presented in Section~\ref{sec:experiments}. Section~\ref{sec:conclusion} concludes the work.

\section{Online CTF Identification}\label{sec:ci}
 
We consider a system with $I$ channels  and one speech source. In the time domain, the $i$-th  microphone signal $x_{i}(n)$ is 
\begin{align}\label{eq:xn}
 x_{i}(n)=s(n)\star a_{i}(n)+e_{i}(n), \quad i=1,\dots,I
\end{align}
where $n$ is the time index, $\star$ denotes convolution, $s(n)$  is the speech source signal, and $a_{i}(n)$ is the RIR from the speech source to the $i$-th microphone. The additive noise term $e_{i}(n)$ will be discarded in the following, since we do not consider noise in this work. In the STFT domain, based on the CTF approximation, we have
\begin{equation}
\label{eq:xpk3}
 x_{i,p,k} \approx s_{p,k}\star a_{i,p,k}, \quad i=1,\dots,I
 \end{equation}
 where $x_{i,p,k}$ and $s_{p,k}$ are the STFT coefficients of the corresponding signals, and the CTF $a_{i,p,k}$ is the subband representation of the RIR $a_{i}(n)$. $p=1,\dots,P$ denotes the STFT frame index and $k=0,\dots,N-1$ denotes the frequency index, $P$ is the number of signal frames in a given processed speech sequence, and $N$ is the STFT frame (window) length. The convolution is executed along the frame index $p$. The length of the CTF, denoted as $Q$, is assumed to be identical for all frequency bins and is approximately equal to the length of the corresponding RIR divided by $L$, where $L$ denotes the STFT frame shift. 

\subsection{Batch CTF Identification}
\addnote[cpx2]{1}{In \cite{li2018taslp}, we proposed a batch mode CTF identification method in the complex domain}.
This method is based on the following cross-relation between channels \cite{xu1995}:
\begin{align}\label{eq:xyha}
 x_{i,p,k}\star a_{j,p,k}=s_{p,k}\star a_{i,p,k}\star a_{j,p,k}=x_{j,p,k}\star a_{i,p,k}. 
\end{align}
However, this equation cannot be directly used. The reason is that, for the  oversampling case (i.e. $L<N$), there is a common region with  magnitude close to zero in the frequency response of the CTFs for all channels, caused by the non-flat frequency response of the STFT window. This common zero frequency region is problematic for the cross-relation method. 
It can be alleviated by using critical sampling (i.e. $L=N$), which however leads to a severe frequency aliasing of the signals. To achieve a good trade-off, it was proposed in \cite{li2018taslp} that the signal STFT coefficients are oversampled to avoid frequency aliasing, but the  multichannel CTF coefficients are forced to be critically sampled to avoid the common zero problem. More precisely, the Hamming window\footnote{Other commonly used windows, such as Hanning and Sine windows, are also applicable.} is used, and we set $L=N/4$ and $L_f=N$, where $L_f$ denotes the frame step of CTF. 
Since the channel identification algorithm presented in this section and the inverse filtering algorithm presented in the next section are both applied frequency-wise, hereafter the frequency index $k$ will be omitted for clarity of presentation. 

\addnote[vecfirst]{1}{Based on the oversampled CTF $a_{i,p}$, the critically sampled CTF is defined in vector form as $\tilde{\mathbf{a}}_i = [a_{i,0},a_{i,4},\dots,a_{i,4(\tilde{Q} -1)}]^{\top}$, where $^{\top}$ denotes matrix/vector transpose and $\tilde{Q} = \lceil Q/4 \rceil$ ($\lceil \cdot \rceil$ is the ceiling function)}. \addnote[cs]{1}{In accordance with this critically sampled CTF, (\ref{eq:xpk3}) should be reformulated with critically sampled source STFT coefficients. However, such reformulation of (2) is actually not used. Instead, in the following CTF identification and inverse filtering methods,  the filtering process is applied to the microphone signals, thence the  STFT coefficients of microphone signals will be critically sampled}. From the oversampled STFT coefficients of microphone signals, we define the convolution vector as ${\tilde{\mathbf{x}}}_{i,p} = [x_{i,p},x_{i,p-4},\dots,x_{i,p-4(\tilde{Q} -1)}]^{\top}, \ p=1,\dots,P$. Note that, when $p<4(\tilde{Q}-1)+1$, the vector ${\tilde{\mathbf{x}}}_{i,p}$ is constructed by padding zeros. Then, the cross-relation  can be recast as 
 \begin{align}\label{eq:cr}
{\tilde{\mathbf{x}}}_{i,p}^{\top} \tilde{\mathbf{a}}_j ={\tilde{\mathbf{x}}}_{j,p}^{\top} \tilde{\mathbf{a}}_i.
\end{align} 
 This convolution formulation can be interpreted as that $3/4$ of the original oversampled CTF coefficients are forced to be zero.
This cross-relation is defined for each microphone pair. To present the cross-relation equation in terms of the CTF of all channels, i.e. 
\begin{align}
\tilde{\mathbf{a}}=[\tilde{\mathbf{a}}^{\top}_1,\tilde{\mathbf{a}}^{\top}_2,\dots,\tilde{\mathbf{a}}^{\top}_I]\tp,
\end{align}
we define: 
 \begin{align}\label{eq:xij}
\tilde{\mathbf{x}}_{ij,p} = [\underbrace{{0},\dots,{0}}_{(i-1)\tilde{Q}},  \tilde{\mathbf{x}}_{j,p}^{\top}, \underbrace{{0},\dots,{0}}_{(j-i-1)\tilde{Q}}, -\tilde{\mathbf{x}}_{i,p}^{\top}, \underbrace{{0},\dots,{0}}_{(I-j)\tilde{Q}}]\tp, \ j>i. 
 \end{align}   
 \addnote[cc]{1}{Then the cross-relation (\ref{eq:cr}) can be written as:}
 \begin{align}\label{eq:xija}
 \tilde{\mathbf{x}}_{ij,p}^{\top} \tilde{\mathbf{a}} = 0.
 \end{align}
 
 There is a total of $M=I(I-1)/2$ distinct microphone pairs, indexed by $(i,j)$ with $j>i$.
 For notational convenience, let $m=1,\dots,M$ denote the microphone-pair index. Then let the subscript $_{ij}$ be replaced with $_m$. 
For the static speaker case, the CTF $\tilde{\mathbf{a}}$ is time-invariant, and can be estimated by solving the following constrained least square problem in batch mode: 
\begin{align}\label{eq:ls1}
 \text{min}  \sum_{p=1}^{P}\sum_{m=1}^{M} |\tilde{\mathbf{x}}_{m,p}^{\top} \tilde{\mathbf{a}}|^2 \quad \text{s.t. }   \mathbf{g}^{\top}\tilde{\mathbf{a}}=1, 
\end{align}
where $| \cdot |$ denotes the (entry-wise) absolute value, and $\mathbf{g}$ is a constant vector 
\begin{align}\label{eq:g}
\mathbf{g} = [1, \underbrace{0, \dots,0}_{\tilde{Q}-1},1, \underbrace{0, \dots,0}_{\tilde{Q}-1}, \dots, 1, \underbrace{0, \dots,0}_{\tilde{Q}-1}]^{\top}.
\end{align}
Here we constrain the sum of the first entries of the $I$ CTFs to be equal to 1, i.e. $\sum_{i=1}^I a^i_0=1$. As discussed in \cite{li2018taslp}, in contrast to the eigendecomposition method proposed in \cite{xu1995}, this contrained least square method is robust against noise interference. The solution to~(\ref{eq:ls1}) is
\begin{align}\label{eq:solution}
\breve{\mathbf{a}}=\frac{\mathbf{R}^{-1}\mathbf{g}}{\mathbf{g}^{\top}\mathbf{R}^{-1}\mathbf{g}},
\end{align}
where $\mathbf{R}$ is the sample covariance matrix of the microphone signals, i.e. $\mathbf{R}=\sum_{p=1}^{P}\sum_{m=1}^{M}\tilde{\mathbf{x}}_{m,p}^{*}\tilde{\mathbf{x}}_{m,p}^{\top}$. 

\subsection{Recursive CTF Identification}
In dynamic scenarios, the CTF vector $\tilde{\mathbf{a}}$ is time-varying, is thus rewritten as $\tilde{\mathbf{a}}^{(p)}$ to specify the frame-dependency. Note that we need to distinguish the superscript $^{(p)}$, which represents the time index with respect to the online update, from the subscript $_p$, which represents the frame index of the signals and filters. At frame $p$, $\tilde{\mathbf{a}}^{(p)}$ can be caculated by (\ref{eq:solution}) using the microphone signals at frame $p$ and recent frames. However, this requires a large amount of inverse matrix calculations, which is computationally expensive. In this work, we adopt the RLS-like algorithm for recursive CTF identification. At the current frame $p$, RLS aims to solve the minimization problem
 \begin{align}\label{eq:rls}
\text{min} \sum_{p'=1}^p \lambda^{p-p'} \Big(\sum_{m=1}^M  |\tilde{\mathbf{x}}_{m,p'}^{\top} \tilde{\mathbf{a}}^{(p)}|^2\Big) \quad \text{s.t. }   \mathbf{g}^{\top}\tilde{\mathbf{a}}=1.
 \end{align} 
The forgetting factor $\lambda^{p-p'}$ with $\lambda \in (0,1]$ gives exponentially decaying weight to older frames. This time-weighted minimization problem can be solved using (\ref{eq:solution}) with $\mathbf{R}$ replaced with a frame-dependent sample covariance matrix $\mathbf{R}^{(p)}=\sum_{p'=1}^p\lambda^{p-p'}(\sum_{m=1}^M  \tilde{\mathbf{x}}_{m,p'}^{*}\tilde{\mathbf{x}}_{m,p'}^{\top})$, namely
\begin{align}\label{eq:psolution}
\breve{\mathbf{a}}^{(p)}=\frac{(\mathbf{R}^{(p)})^{-1}\mathbf{g}}{\mathbf{g}^{\top}(\mathbf{R}^{(p)})^{-1}\mathbf{g}}.
\end{align}
$\mathbf{R}^{(p)}$ can be recursively updated as 
\begin{align}\label{eq:rp}
\mathbf{R}^{(p)} = \lambda\mathbf{R}^{(p-1)}+\sum_{m=1}^M \tilde{\mathbf{x}}_{m,p}^{*}\tilde{\mathbf{x}}_{m,p}^{\top}.  
\end{align}
The covariance matrix is updated in $M$ steps, where each step modifies the covariance matrix by adding a rank-one matrix $\tilde{\mathbf{x}}_{m,p}^{*}\tilde{\mathbf{x}}_{m,p}^{\top}$, $m=1,\dots,M$. To avoid  explicit inverse matrix computation, instead of $\mathbf{R}^{(p)}$ itself, we recursively estimate its inverse $(\mathbf{R}^{(p)})^{-1}$ based on the Sherman-Morrison
formula (\ref{eq:sm}).  This procedure is summarized in Algorithm~\ref{alg:sm}, where the Sherman-Morrison formula is applied in each of $M$ loops.  As an  initialization, we set $(\mathbf{R}^{(0)})^{-1}$ to $1, 000\mathbf{I}$, where $\mathbf{I}$ denotes identity matrix. \addnote[algpw]{1}{The computational complexity of Algorithm~\ref{alg:sm} is proportional to the squared number of microphones. It is found by experiments that the microphone pairs are actually highly redundant for CTF estimation. Therefore, in practice, only the $I-1$ microphone pairs that involve one specific microphone, e.g. the first microphone, are used. This achieves similar performance with using all microphone pairs.}  

 \begin{algorithm}[h]  \caption{\label{alg:sm} Recursive estimation of $(\mathbf{R}^{(p)})^{-1}$ at frame $p$} 
\begin{algorithmic} 
 \STATE Inputs: $\tilde{\mathbf{x}}_{m,p}$, $m=1,\dots,M$; $(\mathbf{R}^{(p-1)})^{-1}$ 
 \STATE Initialization: $\mathbf{P}\leftarrow\lambda^{-1}(\mathbf{R}^{(p-1)})^{-1} $ 
 \FOR{each microphone pair $m=1$ to $M$} 
 \vspace{-0.3cm}
 \STATE \begin{equation}\label{eq:sm}
 \mathbf{P}\leftarrow \mathbf{P}- (\mathbf{P}\tilde{\mathbf{x}}_{m,p}^{*}\tilde{\mathbf{x}}_{m,p}^{\top}\mathbf{P})/(1+\tilde{\mathbf{x}}_{m,p}^{\top}\mathbf{P}\tilde{\mathbf{x}}_{m,p}^{*})
 \end{equation}  
 \ENDFOR
 \STATE Output: $(\mathbf{R}^{(p)})^{-1}\leftarrow\mathbf{P}$ 
  \end{algorithmic}
\end{algorithm}

The number of frames used to estimate $\tilde{\mathbf{a}}^{(p)}$ should be proportional to the length of the critically sampled CTF, i.e. $\tilde{Q}$, and is thus denoted with $\tilde{P}=\rho\tilde{Q}$. On the one hand, a large $\tilde{P}$ is required to ensure estimation accuracy. On the other hand, $\tilde{P}$ should be set as small as possible to reduce the dependency of the estimation on the past frames, namely to reduce the latency of the estimation, which is especially important for the moving speaker case. Similar to the RIR samples, the critically sampled CTF coefficients can be assumed to be temporally uncorrelated. However, the microphone signals STFT coefficients are highly correlated due to the temporal correlation of time-domain speech samples and to the oversampling of signals STFT coefficients (i.e. large overlapping of STFT frames). Empirically, we set $\rho=2.5\times4=10$, where the factor $4$ is used to compensate the signal oversampling effect. To approximately have a memory of $\tilde{P}$ frames, we can set $\lambda=\frac{\tilde{P}-1}{\tilde{P}+1}$.

\section{Adaptive STFT Magnitude Inverse Filtering}\label{sec:dereverberation}

In \cite{li2018taslp}, it was found that the estimated complex-valued CTF  is not accurate enough for effective inverse filtering, due to the influence of noise interference and the frequency aliasing caused by critical sampling. To reduce the sensitivity of the inverse filtering procedure to the CTF perturbations, instead of the complex-valued CTF convolution (\ref{eq:xpk3}), its magnitude approximation was used, i.e. 
\begin{align}\label{eq:magconv}
|x_{i,p}| \approx |s_{p}|\star |a_{i,p}|, \quad i=1,\dots,I.
\end{align}
This magnitude convolution model is widely used in the context of dereverberation, e.g. \cite{mirsamadi2014,mohammadiha2016,baby2017}. In \cite{li2018icassp,li2018taslp2}, we proposed a MINT method based on the complex-valued CTF convolution for multisource separation and dereverbation. In the present work, we adapt this MINT method to the magnitude domain, and develop its adaptive  version for online dereverbation. 

\subsection{Adaptive MINT in the Magnitude Domain}\label{sec:amint}

The CTF estimate of each channel, denoted by $\breve{\mathbf{a}}_i^{(p)}$, $i=1,\dots,I$, can be extracted from $\breve{\mathbf{a}}^{(p)}$. Let $\bar{\mathbf{a}}_i^{(p)}=|\breve{\mathbf{a}}_i^{(p)}|$ denote the CTF magnitude vector, and $\bar{a}_{i,0}^{(p)},\dots,\bar{a}_{i,\tilde{Q}-1}^{(p)}$ its elements.
Define the inverse filters of $\bar{\mathbf{a}}_i^{(p)}$  
in vector form  as $\mathbf{h}_i^{(p)}\in\mathbb{R}^{\tilde{O}\times1}$, $i=1,\dots,I$, where $\tilde{O}$ is the length of the inverse filters, which is assumed to be identical for all channels. Note that both $\bar{\mathbf{a}}_i^{(p)}$ and $\mathbf{h}_i^{(p)}$ are critically sampled. To apply the magnitude inverse filtering using $\mathbf{h}_i^{(p)}$, we construct the STFT magnitude vector of microphone signals as ${\bar{\mathbf{x}}}_{i,p} = [|x_{i,p}|,|x_{i,p-4}|,\dots,|x_{i,p-4(\tilde{O}-1)}|]^{\top}$. The output of the multichannel inverse filtering is given by 
\begin{align}\label{eq:inv}
{\bar{s}}_p = \sum_{i=1}^I\mathbf{h}_i^{(p) \top} \bar{\mathbf{x}}_{i,p}.
\end{align}
This output should target the STFT magnitude of the source signal, i.e. $|s_p|$. 

To this aim, the multichannel equalization, i.e. MINT, should target an impulse function, namely
\begin{align}\label{eq:mint1}
\sum_{i=1}^I\bar{\mathbf{A}}_i^{(p)}\mathbf{h}_i^{(p)} = \mathbf{d},
\end{align} 
where the impulse function $\mathbf{d}$ is defined by $\mathbf{d}=[1,0,\dots,0]^{\tp}\in\mathbb{R}^{(\tilde{Q}+\tilde{O}-1)\times1}$, and the convolution matrix $\bar{\mathbf{A}}_i^{(p)}$ is defined by  
\begin{equation}\label{eq:cm}
\bar{\mathbf{A}}_i^{(p)} =
\begin{bmatrix}
\bar{a}_{i,0}^{(p)} & 0     &  \cdots   & 0  \\
\bar{a}_{i,1}^{(p)} & \bar{a}_{i,0}^{(p)} &  \ddots  & \vdots     \\
\vdots & \ddots &   \ddots & \vdots     \\ 
\bar{a}_{i,\tilde{Q}-1}^{(p)} & \vdots & \ddots & 0   \\ 
0 & \bar{a}_{i,\tilde{Q}-1}^{(p)} & \ddots & \vdots \\
\vdots & \ddots & \ddots    & \vdots \\
0& \cdots &0  & \bar{a}_{i,\tilde{Q}-1}^{(p)} \\
\end{bmatrix} \in \mathbb{R}_{\ge 0}^{(\tilde{Q}+\tilde{O}-1)\times \tilde{O}}.   
\end{equation}
In a more compact form, we can write
\begin{align}\label{eq:mint}
\bar{\mathbf{A}}^{(p)}\mathbf{h}^{(p)}=\mathbf{d},
\end{align}
where $\bar{\mathbf{A}}^{(p)}=[\bar{\mathbf{A}}_1^{(p)},\dots,\bar{\mathbf{A}}_I^{(p)}]\in\mathbb{R}_{\ge 0}^{(\tilde{Q}+\tilde{O}-1)\times I\tilde{O}}$ and $\mathbf{h}^{(p)}=[\mathbf{h}_1^{(p)\top},\dots,\mathbf{h}_I^{(p) \top}]^{\top}\in\mathbb{R}^{I\tilde{O}\times1}$. The inverse filter estimation amounts to solving problem (\ref{eq:mint}), or equivalently minimizing the squared error 
\begin{align}\label{eq:ls}
J^{(p)} = \parallel \bar{\mathbf{A}}^{(p)}\mathbf{h}^{(p)}-\mathbf{d} \parallel ^2,
\end{align}
where $\parallel\cdot \parallel$ denotes $\ell_2$-norm. The size of $\bar{\mathbf{A}}^{(p)}$ can be adjusted by tuning the length of the inverse filter, i.e. $\tilde{O}$. If $\bar{\mathbf{A}}^{(p)}$ is square or wide, i.e. $(\tilde{Q}+\tilde{O}-1)\le I\tilde{O}$ and thus $\tilde{O}\ge\frac{\tilde{Q}-1}{I-1}$, (\ref{eq:mint}) has an exact solution and (\ref{eq:ls}) can reach zero. Otherwise, (\ref{eq:mint}) is a least square problem, and only an approximate solution can be achieved. 

The minimization of (\ref{eq:ls}) has a closed-form solution.  
However, this needs the computation of an inverse matrix for each frame and frequency, which is computationally expensive. In this work, we propose to adaptively estimate $\mathbf{h}^{(p)}$ following the principle of normalized LMS. For a summary of normalized LMS design and analysis, please refer to Chapter 10.4 of \cite{manolakis2000}. The proposed LMS-like adaptive estimation method presented in the following is based on a stationary filtering system, but can be directly used for the nonstationary case due to its natural adaptive characteristic. In a stationary system, the filter to be estimated, i.e. the inverse filter  $\mathbf{h}$ in the present work, is assumed to be time-invariant. The aim of LMS is to adaptively minimize the mean squared error $\mathbb{E}[J]$, where $\mathbb{E}[\cdot]$ denotes expectation. Note that with the superscript $^{(p)}$ removed, $\mathbf{h}$ and ${J}$ denote the stationary filter and the (stationary) random variable for the squared error, respectively.
At frame $p$, the instantaneous filtering process in (\ref{eq:mint}) and the squared error (\ref{eq:ls}) are a random instance of the stationary system.   At frame $p$, the adaptive update uses the gradient of the instantaneous error $J^{(p)}$ at the previous estimation point $\mathbf{h}^{(p-1)}$, i.e.
\begin{align}\label{eq:gra}
\Delta J^{(p)}|_{\mathbf{h}^{(p-1)}} =  2\bar{\mathbf{A}}^{(p) \top}(\bar{\mathbf{A}}^{(p)}\mathbf{h}^{(p-1)}-\mathbf{d}).
\end{align}
An estimate of $\mathbf{h}^{(p)}$ based on the gradient descent update is
\begin{align}\label{eq:grades}
\mathbf{h}^{(p)} = \mathbf{h}^{(p-1)}-\frac{\mu}{\text{Tr}(\bar{\mathbf{A}}^{(p) \top}\bar{\mathbf{A}}^{(p)})}\Delta J^{(p)}|_{\mathbf{h}^{(p-1)}}, 
\end{align}
where $\text{Tr}(\cdot)$ denotes the matrix trace,  and $\frac{\mu}{\text{Tr}(\bar{\mathbf{A}}^{(p) \top}\bar{\mathbf{A}}^{(p)})}$ is the \emph{step-size} for gradient descent. The normalization term $\frac{1}{\text{Tr}(\bar{\mathbf{A}}^{(p) \top}\bar{\mathbf{A}}^{(p)})}$ is set to make the gradient descent update converge to an optimal solution, namely to ensure the update stability.  It is proven in \cite{manolakis2000} that, to guarantee the stability, the \emph{step-size} should be set to be lower than $\frac{1}{\text{Tr}(\mathbb{E}[\bar{\mathbf{A}}^{\top}\bar{\mathbf{A}}])}$, where $\bar{\mathbf{A}}$ denotes the (stationary) random variable for the CTF convolution matrix. Following the principle of normalized LMS, we replace the expectation $\mathbb{E}[\bar{\mathbf{A}}^{\top}\bar{\mathbf{A}}]$ with the instantaneous matrix $\bar{\mathbf{A}}^{(p) \top}\bar{\mathbf{A}}^{(p)}$. The matrix trace can be computed as $\text{Tr}(\bar{\mathbf{A}}^{(p) \top}\bar{\mathbf{A}}^{(p)})=\tilde{Q}\sum_{i=1}^I \bar{\mathbf{a}}_i^{(p) \top}\bar{\mathbf{a}}_i^{(p)}$. The constant step factor $\mu$ ($0<\mu\le 1$) should be empirically set to  achieve a good tradeoff between convergence rate (and tracking ability in a dynamic scenarios with time-varying CTFs) and update stability. 

The proposed magnitude inverse filtering method is summarized in Algorithm~\ref{alg:invfil}, which is recursively executed frame by frame.  As an initialization, we set $\mathbf{h}^{(0)}$ to a vector with all entries being zero.

\begin{algorithm}[t]  \caption{\label{alg:invfil} Adaptive STFT magnitude inverse filtering at frame $p$} 
\begin{algorithmic} 
 \STATE Input: $\breve{\mathbf{a}}^{(p)}$ computed by~(\ref{eq:psolution}) and $\mathbf{h}^{(p-1)}$. 
 \STATE 1 Construct $\bar{\mathbf{A}}^{(p)}$ using~(\ref{eq:cm}),
 \STATE 2 Compute gradient $\Delta J^{(p)}|_{\mathbf{h}^{(p-1)}}$ using~(\ref{eq:gra}),
 \STATE 3 Update inverse filter $\mathbf{h}^{(p)}$ using~(\ref{eq:grades}),
 \STATE 4 Estimate the speech signal STFT magnitude ${\bar{s}}_p$ with inverse filtering (16).   
 \STATE Output: ${\bar{s}}_p$ and $\mathbf{h}^{(p)}$.
  \end{algorithmic}
\end{algorithm}

\subsection{Multichannel Processing}\label{sec:mp}


In the time-domain and complex-valued CTF MINT methods, e.g., \cite{kodrasi2013,li2018icassp,lim2014}, the optimal inverse filtering performance is achieved by setting the length of the inverse filter to the smallest value that makes $\bar{\mathbf{A}}^{(p)}$ be square or slightly wide, i.e. $\tilde{O}=\lceil \frac{\tilde{Q}-1}{I-1} \rceil$. This means $\tilde{O}$ becomes smaller with the increase of the number of channels.
However, for the present magnitude inverse filtering method, this configuration is only suitable  for the two-channel case. 
For the two-channel case, the length of the inverse filters $\tilde{O}=\tilde{Q}-1$ is close to the CTF length $\tilde{Q}$, and in our experiments we actually set $\tilde{O}=\tilde{Q}$. The STFT magnitude of the microphone signals for the current frame includes the information of the past $\tilde{Q}-1$ frames of the speech source signal due to the CTF convolution. Therefore, it is reasonable that the magnitude inverse filtering at the current frame uses the past $\tilde{Q}-1$ frames of the microphone signals to remove the reflections..  
When the number of channels is larger than two, the configuration $\tilde{O}=\lceil \frac{\tilde{Q}-1}{I-1} \rceil$ leads to a very small $\tilde{O}$, since the length of the critically sampled CTF, i.e. $\tilde{Q}$, is already relatively small. As will be shown in the experiments section, $\tilde{Q}$ is related to both the STFT setting and the reverberation time, and is set to $4$ in this work. 
\addnote[iflength]{1}{For the time-domain and complex-valued CTF MINT methods \cite{kodrasi2013,li2018icassp,lim2014}, dereverberation is guaranteed by solving the multichannel MINT equation, regardless of the length of the inverse filter, since the time-domain and CTF convolutions are exactly evaluated. By contrast, the magnitude convolution (\ref{eq:magconv}) is a rough approximation.  Even if the magnitude MINT (\ref{eq:mint}) can be exactly solved with a very small $\tilde{O}$, experiments show that the resulting magnitude inverse filtering is not able to efficiently suppress reverberation.} 

As detailed below, we propose two multichannel processing schemes suitable for the present magnitude inverse filtering method. They are both evaluated in Section~\ref{sec:experiments}.
\addnote[mcmint]{1}{\subsubsection{Multichannel magnitude MINT with $\tilde{O}=\tilde{Q}$ regardless of the number of channels} This exactly follows the formulations presented in Section \ref{sec:amint}. The setting $\tilde{O}=\tilde{Q}$ is motivated by the principle that, as is done for the two-channel case, the reflection magnitude of the past $\tilde{Q}-1$ frames should be subtracted from the magnitude of the current frame.} 
\subsubsection{Pairwise magnitude MINT} First, the adaptive MINT (and inverse filtering) presented in Section~\ref{sec:amint} is separately applied for each microphone pair. 
Then the estimates of the source magnitude obtained by all the $M$ microphone pairs are averaged as a new source magnitude estimate, which is still denoted by ${\bar{s}}_p$ for brevity. The source magnitude estimates provided by the different microphone pairs are assumed to be independent, thence the average of them is hopefully suffering from lower interferences and distortions than each of them. 


\subsection{Postprocessing}

The above STFT magnitude inverse filtering does not automatically guarantee the non-negativity of ${\bar{s}}_p$, which is infeasible solution for the STFT magnitude of the source signal. Negative values generally appear for the microphone signal frames with a magnitude that is considerably smaller than the magnitude in the preceding frames. Indeed, in that case, applying negative inverse filter coefficients to the preceding frames produces a negative magnitude estimate. Such frames are normally  following a high-energy speech region, but themselves include very low source energy or purely reverberations. To overcome this, one way is to add the non-negativity constraint of the inverse filtering output to (\ref{eq:ls}), which however leads to a larger complexity for both algorithm design and computation. Instead, we constrain the lower limit of the STFT magnitude of source signal according to the (averaged) STFT magnitude of microphone signals. Formally, the final estimate of the STFT magnitude of source signal is
\begin{align}\label{eq:limit}
\check{s}_p = \text{max}({\bar{s}}_p,G_{\text{min}}\frac{1}{I}\sum_{i=1}^I|x_{i,p}|),
\end{align}
where $G_{\text{min}}$ is a constant lower limit gain factor. This type of lower limit is widely used in single-channel speech enhancement methods, e.g. in \cite{cohen2001}, mainly to keep the noise naturalness. \addnote[nncons]{1}{In the experiments described below, about 20\% of TF bins are modified by this constraint.}

Finally, the STFT phase of one of the microphone signals, e.g. the first microphone is used in this work, is taken as the phase of the estimated STFT coefficient of source signal, i.e. we have $\hat{s}_{p}=\check{s}_pe^{j\arg[x^1_{p}]}$, where $\arg[\cdot]$ is the phase of complex number.
The time-domain source signal $\hat{s}(n)$ is obtained by applying the inverse STFT. Note that the MINT formulation (\ref{eq:mint}) implies that the proposed inverse filtering method aims at recovering the signal corresponding to the first CTF frame, which not only includes the direct-path impulse response, but also the early reflections within the duration of one STFT frame. As a result, the estimated source signal $\hat{s}(n)$ includes both the direct-path source signal and its early reflections within $N/f_s$ seconds following the direct-path propagation, where $f_s$ is the signal sampling rate. 

\subsection{Difference from Conventional MINT Methods}
Due to the use of i) the magnitude convolution model, ii) the critically sampled CTFs and inverse filters, and iii) the adaptive update of the inverse filters, the present adaptive MINT method is largely different from the complex-valued CTF MINT  \cite{li2018icassp,li2018taslp2} and the time-domain MINT, such as \cite{kallinger2006,kodrasi2013,lim2014,hikichi2007,mertins2010}. 
Besides the pairwise processing scheme, the two main differences are the following.     

\subsubsection{Desired Response of MINT}
In many time-domain methods, to improve the robustness of MINT to microphone noise and filter perturbations, the target function (desired response) is designed to have multiple non-zero taps. This can be done either by explicitly filling the target function with multiple non-zero taps, such as the partial MINT in \cite{kodrasi2013}, or by relaxing the constraint for some taps, such as the relaxed multichannel least-squares in \cite{lim2014}. This way, the desired response with multiple non-zero taps includes both the direct-path propagation and some early reflections. In the present work, the impulse function $\mathbf{d}$ is used as the desired response of MINT in the CTF domain, namely only one non-zero tap is sufficient, since one tap of CTF corresponds to a segment of RIR that includes both direct-path propagation and early reflections. 
 
It was shown in \cite{li2018icassp,li2018taslp2} that, due to the effect of short time STFT windows, the oversampled CTF of multiple channels have common zeros, which is problematic for MINT. A target function incorporating the information of the STFT windows was proposed to compensate the common zeros.  In the present work, the critically sampled CTFs do not suffer from this problem.       

A modeling delay is always used in the time-domain MINT and complex-valued CTF MINT methods, i.e., in the target function, a number of zeros are inserted prior to the first non-zero tap. It is shown in \cite{li2018icassp,hikichi2007} that the optimal length of the modeling delay is related to the direct-path tap and the length of the room filters.    
In the present method, the room filters, i.e. CTFs, are blindly estimated, with the direct-path lying in the first tap. In addition, the CTF length is very small as mentioned above. Therefore, the modeling delay is set to 0, which achieved the best performance in our experiments. 
 
\subsubsection{Energy Regularization}
An energy regularization is used in \cite{kodrasi2013,li2018icassp,hikichi2007} to limit the energy of the inverse filters derived by MINT, since high energy inverse filters  will amplify microphone noise and filter perturbations. For example, in the present problem, the optimal MINT solution could have a very large energy, especially when the matrix $\bar{\mathbf{A}}^{(p)  \top}\bar{\mathbf{A}}^{(p)}$ is ill-conditioned. However, for the proposed method, the inverse filters are adaptively updated based on the previous estimation. The step size is set with  guaranteed update stability. Thence, the energy of the inverse filters will not be boosted once the inverse filters are properly initialized.

\section{Experiments} \label{sec:experiments}

\subsection{Experimental Configuration}

\subsubsection{Dataset}
We evaluate the proposed method using two datasets.
\paragraph{The REVERB challenge dataset \cite{kinoshita2016}} We used the evaluation set  of SimData-room3 and RealData datasets. SimData-room3 was generated by convolving clean signals from the WSJCAM0 dataset with RIRs measured in a room with  reverberation time $T_{60}=$ 0.7 s, and adding  pre-recorded stationary ambient noise with  an SNR of 20 dB. The microphone-to-speaker distances are 1~m (\emph{near}) and 2~m (\emph{far}). \addnote[drr1]{1}{For these two distances, the \textit{direct-to-reverberation ratios} (DRRs)  are 10.6 dB and 1.0 dB, respectively, and the \textit{early-to-late reverberation ratios} ($C_{50}$) are 14.9 dB and 6.3 dB, respectively.}  RealData was recorded  in a noisy room  with $T_{60}=0.7$~s (different room than \emph{SimData-room3}) and where humans pronounce MC-WSJ-AV utterances \cite{lincoln2005} microphone-to-speaker distances of 1~m (\emph{near}) and 2.5~m (\emph{far}). We used the data captured with two microphones (2-ch) or an eight-channel circular microphone array (8-ch). 

We tested the automatic speech recognition (ASR) performance obtained with the enhanced signals, in addition to the speech enhancement performance. \addnote[kaldi]{1}{The ASR system provided by \cite{weninger2014,peddinti2015}, with the Kaldi recipe,\footnote{https://github.com/kaldi-asr/kaldi/tree/master/egs/reverb} is taken as the baseline system. This system uses  Mel-frequency cepstral coefficients (MFCC) and iVector \cite{dehak2011} features, time-delay neural network (TDNN) acoustic model, and the WSJ 5k vocabulary and trigram language model. TDNN is capable to learn the long-term temporal dynamics of speech signals including the effects of reverberation.} \addnote[traindata]{1}{TDNN is trained using the multi-condition WSJCAM0 training dataset. The eight-channel multi-condition data are generated by convolving the 7,861 utterances of clean WSJCAM0 training signals with real recorded RIRs, and adding pre-recorded stationary ambient noise with an SNR of 20 dB. The eight-channel multi-condition data are then speed-perturbed with speed factors of 0.9, 1 and 1.1. In total, $7,861\times 8\times 3=188,664$ reverberant and speed-perturbed multi-condition utterances are used for TDNN training, which represents a total speech signal duration of about 373 hours}. To account for the online nature of the proposed method, the online ASR decoding  provided in the REVERB Kaldi recipe is used. 
\paragraph{The Dynamic dataset \cite{schwartz2015online}} This dataset was recorded by an eight-channel linear microphone array and a close-talk microphone in a room with $T_{60}=$ 0.75 s. \addnote[drr2]{1}{The average DRR and $C_{50}$ values for this dataset are $-5.5$ dB and $3.0$ dB, respectively}. The recording SNR is about 20 dB. Four human speakers read an article from the New-York Times.  Speakers could be static, or moving slightly, such as when standing up, sitting down and turning their head, or moving largely such as moving from one point to another. Speakers could be facing or not facing the microphone array. The total length of the dataset is 48 minutes. We split the data into three subsets: i) A subset with speakers being static  and facing the microphone array (Static-FA). Note that some slight movements are inevitable even if human speakers are asked to be static; ii)  Static and not facing the array (Static-NFA), and iii) Moving from one point to another. We used the central two channels (2-ch) or all the eight channels (8-ch). \addnote[recognizer]{1}{As for ASR, some pilot experiments show that the REVERB recognizer performs poorly for this dataset, since a number of words in this dataset are not in the  WSJ 5k vocabulary. Instead, we used Google Cloud Speech-to-Text\footnote{https://cloud.google.com/speech-to-text/} to conduct the ASR experiment on this dataset.}

\subsubsection{Parameter Settings}
The following parameter settings are used for both datasets, and all the experimental conditions. The sampling rate is 16 kHz. The STFT uses a Hamming window with length of $N = 768$ (48 ms) and frame step $L = N/4 = 192$ (16 ms). As a result, the 48 ms early reflections will be preserved in the dereverberated signal. It is shown in \cite{sehr2010} that, to achieve a better ASR performance, early reflections should be removed as much as possible when late reverberation is perfectly removed. However, when the remaining late reverberation is not low, ASR performance benefits from preserving more early reflections up to 50 ms. Therefore, as we are dealing with adverse acoustic conditions, such as intense reverberation/noise
or moving speakers, where late reverberation cannot be
perfectly suppressed, we have decided to preserve the early reflections in the first 48 ms. 
The CTF length $Q$ (and $\tilde{Q}$) is related to the reverberation time, and is the only prior knowledge that the proposed method requires. It is set to $Q=16$ (and $\tilde{Q}=4$), which covers the major part of the RIRs, and also excludes a heavy tail. According to the CTF length, the forgetting factor $\lambda$ is set to $\frac{40-1}{40+1}\approx 0.95$. The constant step factor $\mu$ is set to  0.025. The constant lower limit gain factor $G_{\text{min}}$ is set to correspond to $-15$ dB.  These parameters are set to achieve the best ASR performance for the RealData subset of  the REVERB challenge dataset, and are directly used for other experimental conditions.    

\setlength{\tabcolsep}{6.0pt}
\begin{table*}[t]
\centering
\caption{\small SRMR, PESQ and STOI metrics (larger the better) for the REVERB challenge dataset.}
\label{tab:se_reverb}
\begin{tabular}{c l | c  c c  |  c c  c| c  c c  |  c c  c  }  
& &  \multicolumn{6}{c|}{SRMR} & \multicolumn{3}{c|}{PESQ} & \multicolumn{3}{c}{STOI} \\
	    &     &  \multicolumn{3}{c|}{SimData-room3}   & \multicolumn{3}{c|}{RealData}  &  \multicolumn{3}{c|}{SimData-room3}&  \multicolumn{3}{c}{SimData-room3}\\ 
	     \vspace{1mm}
	ch    &  & \emph{near}  & \emph{far}   & Average     & \emph{near}  & \emph{far}  & Average  & \emph{near}  & \emph{far}   & Average& \emph{near}  & \emph{far}   & Average\vspace{1mm} \\ \hline \vspace{1mm}
& unproc.      & 2.35  & 2.29  & 2.32    & 2.29  & 2.20 & 2.24 & 1.89  & 1.55 & 1.72   & 0.89  & 0.71 & 0.80 \\
2-ch	&BWPE        & 2.44  & 2.42  & 2.43    & 2.55  & 2.54 & 2.55 & 2.06 & 1.67 & 1.87  & 0.92  & 0.78 & 0.85    \\
        &AWPE        & 2.61  & 2.84  & 2.73    & 2.99  & 2.98 & 2.99 & 2.32  & 1.77 & 2.05  & 0.78  & 0.76 & 0.77    \\ \vspace{2mm}
        &SMIF (ours)      & 2.51  & 2.63 & 2.57    & 2.83  & 2.76 & 2.80  & 2.25  & 1.74 & 2.00  & 0.77  & 0.73 & 0.75 \\
8-ch	&BWPE        & 2.49  & 2.59 & 2.54    & 2.79  & 2.83 & 2.81	& 2.38 & 2.10 & 2.24 & 0.94  & 0.87 & 0.91\\
        &AWPE        & 2.60  & 2.89 & 2.75    & 3.04  & 3.01 & 3.03	& 2.48  & 1.90 & 2.19 & 0.80  & 0.79 & 0.80\\ 
        &SMIF-MC (ours)   & 2.50  & 2.64 & 2.57    & 2.88  & 2.80 & 2.84 & 2.35  & 1.78 & 2.07	& 0.76  & 0.74 & 0.75\\	
	    &SMIF-PW (ours)   & 2.51  & 2.72 & 2.62    & 2.94  & 2.87 & 2.91 & 2.40  & 1.84 & 2.12	& 0.78  & 0.75 & 0.77\\	
\end{tabular}
\end{table*} 

\subsubsection{Comparison Method} 
We compare the proposed method with the adaptive weighted prediction error (AWPE) method presented in \cite{caroselli2017}. The STFT uses a Hanning window with a length of 512 (32 ms) and frame step of 128 (8 ms). For the 2-ch and 8-ch cases, the length of the linear prediction filters is set to 16 and 8,  respectively. The prediction delay is set to 6 to also involve 48 ms of early reflections in the dereverberated signal.  In RLS, the length of the prediction filter vector to be estimated is equal to the length of the filters times the number of channels. Some pilot experiments show that, to obtain the optimal performance, the number of frames used to estimate the prediction filter vector should be set to be twice the vector length. Accordingly, the forgetting factor in RLS is set to 0.97 and 0.985 for the 2-ch and 8-ch cases, respectively. The first channel is taken as the target channel.  Note that these parameters are also set to achieve the best ASR performance for RealData of REVERB challenge dataset, and are directly used for other experimental conditions. 

To evaluate the effectiveness of the online realization of AWPE and the proposed method, we also conducted experiments using these methods implemented in offline (batch) mode. \textbf{i)} \addnote[bwpe]{1}{For the REVERB challenge dataset, the offline WPE is tested. We used the  Python software package \cite{drude2018}, which is integrated in the REVERB kaldi recipe. We adopted the WPE parameters as set by the authors of REVERB kaldi recipe, which are supposed to have been optimally tuned. The STFT configuration was the same as our AWPE implementation, namely using Hanning window with a length of 512 and a frame step of 128. The prediction delay is set to 3.  The length of the linear prediction filters was set to 10 for both the 2-ch and 8-ch cases. The number of iterations for speech variance estimation was set to 5. We refer to this offline WPE as BWPE (batch WPE)}. \textbf{ii)} \addnote[batch]{1}{For the Dynamic dataset, the batch mode counterpart of the proposed method was tested. The CTF identification was conducted in batch mode using (\ref{eq:solution}). Since the magnitude MINT in batch mode has not been investigated, we used the adaptive magnitude MINT presented in Section \ref{sec:dereverberation} for inverse filtering, where the inverse filter $\mathbf{h}^{(p)}$ quickly converged to a constant due to the use of the constant offline estimated CTF.}

\subsubsection{Performance Metrics}
To evaluate the speech enhancement performance, three measures are used, i) a non-intrusive metric, i.e. normalized speech-to-reverberation modulation energy ratio (SRMR) \cite{santos2014improved}, which mainly measures the amount of reverberation and noise, and also reflects the speech intelligibility; and two intrusive metrics ii) perceptual evaluation of speech quality (PESQ) \cite{rix2001}  evaluates the quality of the enhanced signal in terms of both reverberation reduction and speech distortion;  iii) short-time objective intelligibility (STOI) \cite{taal2011} is a metric that highly correlates with speech intelligibility.
To measure PESQ and STOI, the clean source signal is taken as the reference signal. For the Dynamic dataset, the close-talk recording is taken as the source signal. For RealData of the REVERB challenge dataset, the clean signals are not available, thus neither PESQ nor STOI metrics are reported in this case. For  these three metrics, the larger the better. 
The ASR performance is measured with the percentage of word error rate (WER): the lower the better. 
Note that all the tested methods do not perform noise reduction, thence the outputs used to calculate the metrics may contain some amount of noise.

\subsection{Results for the REVERB Challenge Dataset}

In the REVERB challenge dataset, each subset involves several hundreds of individual signals, with each signal being one utterance spoken by one static speaker. The relative speaker-microphone position changes from utterance to utterance. To simulate a realistic turn-taking scenario, for each subset, all the individual signals are first concatenated as a long signal, which is then processed by the online dereverberation methods, i.e. AWPE and the proposed method. The long enhanced signal is finally separated corresponding to the original individual signals. For BWPE, the individual signals are separately processed.  The perfomance measures are computed using the individual enhanced signals. 

\subsubsection{Speech Enhancement Results}
We refer to the proposed method as \textbf{SMIF} (Spectral Magnitude Inverse Filtering). For the multichannel case, the two schemes proposed in Section \ref{sec:mp}, i.e. multichannel processing and pairwise processing, are refered to SMIF-MC and SMIF-PW, respectively.  
Table~\ref{tab:se_reverb} presents the speech enhancement results. As for the proposed method, compared to the 2-ch case, the 8-ch SMIF-MC method improves the SRMR and PESQ metrics on RealData, and achieves identical SRMR and STOI metrics on the SimData-room3 data. The 8-ch SMIF-PW method systematically outperforms the 2-ch case and the 8-ch SMIF-MC method. This indicates that, for the SMIF-MC method, the magnitude inverse filtering accuracy can be improved by using more microphones, however the improvement is not always significant in terms of speech enhancement metrics. In the 8-ch SMIF-PW method, the average of pairwise source estimates successfully suppress the interferences and distortions of the one-pair source estimates. Informal listening tests show that the residual late reverberation can be sometimes noticeably perceived for the 2-ch case, while it is not clearly audible for the 8-ch case. 

\addnote[staticreason]{1}{For all conditions and for all metrics, AWPE outperforms the proposed method, especially the gaps between SRMR metrics are noticeable, see Table~\ref{tab:se_reverb}. The proposed method is based on the STFT-magnitude convolution and inverse filtering, which is a coarse approximation of the real filtering process. By contrast, AWPE is based on a more accurate complex-valued inverse filtering. As a result, the dereverberated signals obtained with the proposed method are likely to have more late reverberation,  extra noise and speech distortions, especially for the 2-ch case}. 
Relative to the unprocessed signal, AWPE and the proposed method slightly improve the STOI metrics for the \emph{far} case, but reduce the STOI metrics for the \emph{near} case. This is possibly because the parameters are set based on the RealData data, and in particular the length of the (inverse) filters may be too large for the \emph{near} simulation data.  

Compared to AWPE, BWPE achieves worse SRMR and 2-ch PESQ metrics, and better 8-ch PESQ and STOI metrics. Generally speaking, BWPE would outperform AWPE if the same parameters were set for both methods, since the speech variance estimate of BWPE is more accurate than the one for AWPE, where the former is iteratively estimated while the latter is approximated by the microphone speech variance. The performance difference between BWPE and AWPE is mainly due to their different prediction delays, i.e. 3 and 6 respectively. A larger prediction delay preserves more early reverberation, which promotes the SRMR metrics, but leads to a larger difference with the clean direct-path signal.

\setlength{\tabcolsep}{4.0pt}
\begin{table}[t]
\centering
\caption{\small WER (\%) for the REVERB challenge dataset.}
\label{tab:wer}
\begin{tabular}{c l | c  c c  |  c c  c  }   
	    &     &  \multicolumn{3}{c|}{SimData-room3}   & \multicolumn{3}{c}{RealData}  \\ 
	     \vspace{1mm}
	  ch  &  & \emph{near}  & \emph{far}   & Average    & \emph{near}  & \emph{far}   & Average  \\ \hline \vspace{1mm}
&unproc.  & 5.08 & 8.08 & 6.58  & 20.95 & 21.27 & 21.11 \\
2-ch	&BWPE        & 4.55  & 6.95 & 5.75  & 15.65  & 15.77 &  15.71    \\
        &AWPE        & 5.37  & 7.28 & 6.33  & 15.36  & 16.21 &  15.79    \\ \vspace{2mm}
        &SMIF (ours)       & 5.01  & 7.16 & 6.09  & 15.30  & 16.04 & 15.67    \\ 
8-ch    &BWPE        & 4.04  & 4.96 & 4.50  & 12.20  & 13.17 & 12.69	\\ 
	    &AWPE        & 4.65  & 6.07 & 5.36  & 12.26  & 13.54 & 12.90	\\    	 
        &SMIF-MC (ours)       & 4.53  & 6.34 & 5.44  & 13.09  & 14.11 & 13.60	 \\
	    &SMIF-PW (ours)       & 4.53  & 5.98 & 5.26  & 13.00  & 14.48 & 13.74	 \\	    
\end{tabular}
\end{table}

\subsubsection{ASR Results}
Table~\ref{tab:wer} presents the WER. It is seen that the present baseline WERs are already very advanced compared with the REVERB challenge WERs reported in \cite{kinoshita2016}. The baseline WERs are noticably reduced by all the tested methods. For instance, as for RealData, the proposed method achieves 25.8\%, 35.6\% and 34.9\% relative WER improvement with 2-ch, 8-ch SMIF-MC and SMIF-PW schemes, respectively. 
In contrast to the speech enhancement metrics presented in Table~\ref{tab:se_reverb}, the ASR performance of the proposed 8-ch SMIF-MC method is noticeably better than the one of the 2-ch case, and is comparable to the one of the 8-ch SMIF-PW method. This means the speech quality improvement caused by the 8-ch SMIF-MC method over the 2-ch case can be well recognized by the ASR system. 

Approximately, the proposed method achieves comparable ASR performance with AWPE. Compared with AWPE, the remaining late reverberation and  extra noise caused by the proposed method  degrades the speech enhancement metrics as shown in Table~\ref{tab:se_reverb}, but can be tackled by the well-trained TDNN acoustic model. 

AWPE does not perform as well as BWPE for SimData-room3, but is comparable to BWPE for RealData. As mentioned above, AWPE preserves more early reflections, which is beneficial for the more challenging RealData, since the late reverberation cannot be well suppressed. Concerning the RealData,  it is possible to further improve the BWPE parameters. However, the parameter tuning for BWPE is out of the scope of this work.


\begin{figure}[t]
\centering
{\includegraphics[width=0.49\columnwidth]{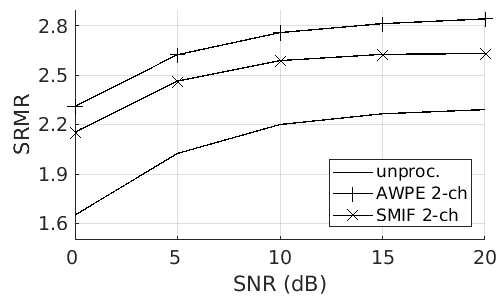}} 
{\includegraphics[width=0.49\columnwidth]{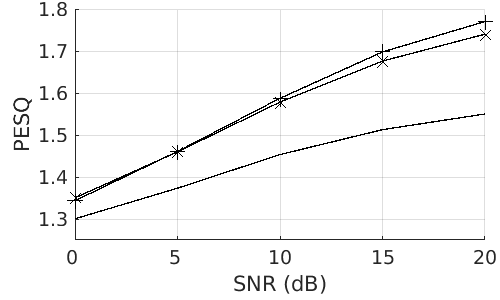}} \\
{\includegraphics[width=0.49\columnwidth]{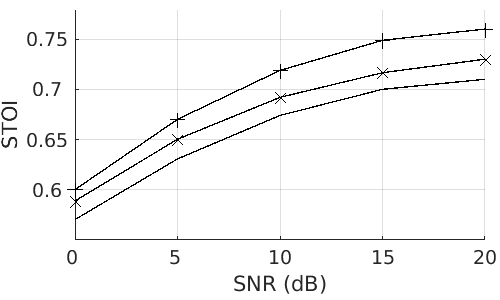}}  
{\includegraphics[width=0.49\columnwidth]{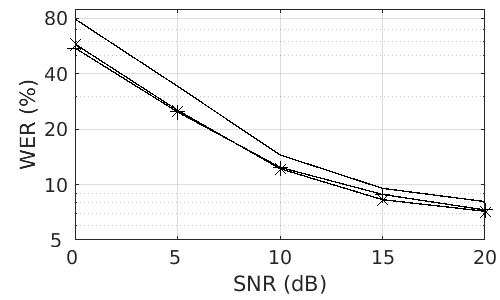}} 
\caption{\small Dereverberation performance as a function of SNR, for the SimData-room3 \emph{far} data.} 
\label{fig:snr}
\end{figure}

\subsubsection{Dereverberation Performance under Noisy Conditions}

To evaluate the sensitivity of the proposed method to noise,  experiments for the SimData-room3 \emph{far} data are conducted with various SNRs. Fig. \ref{fig:snr} shows the results. As expected, the performance of the proposed method decreases with the decrease of SNR, and it has a similar decrease rate with the performance of AWPE. In terms of SRMR, the performance of the two methods have a similar decrease rate with the one of the unprocessed signals, and the performance improvement of the two methods over the unprocessed signals is still significant when SNR is low, e.g. 0 dB. For PESQ and STOI, the performance metrics of the two methods gradually approach the metrics of the unprocessed signals with the decrease of SNR. This means these two metrics are dominated by the intense noise for the low SNR cases. The WER improvement of the two methods over the unprocessed signals are even larger for the low SNR cases than for the high SNR cases. This indicates that reverberation degrades the ASR performance more significantly when it is combined with noise than itself alone, and the two methods are able to efficiently suppress reverberation under intense noise condition.

\subsection{Results for Dynamic Dataset}

\begin{figure}[t]
\centering
{\includegraphics[width=0.9\columnwidth]{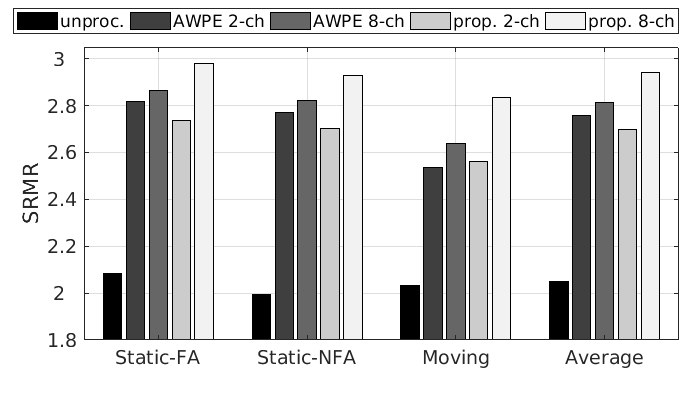}} \\
{\includegraphics[width=0.9\columnwidth]{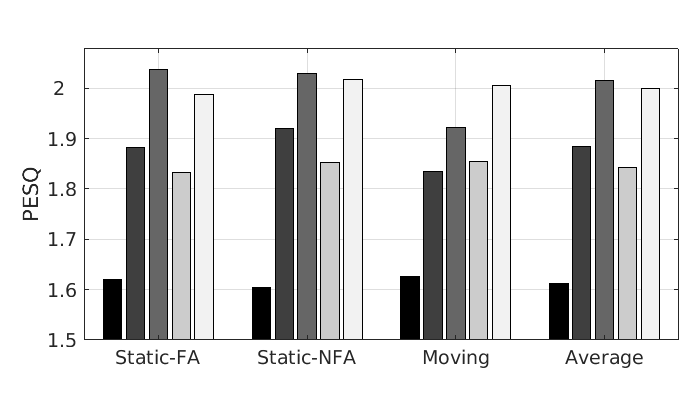}} \\
{\includegraphics[width=0.9\columnwidth]{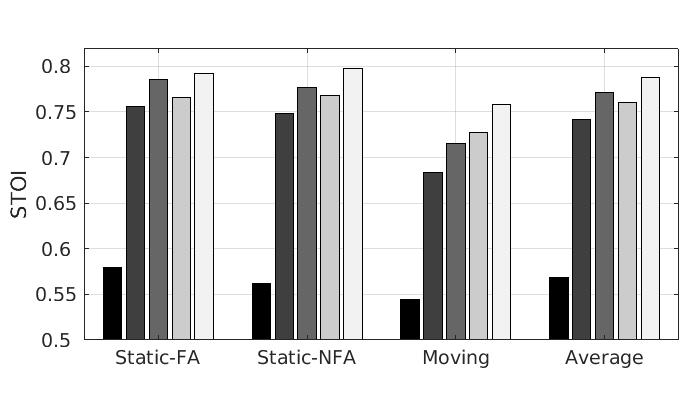}} \\
{\includegraphics[width=0.9\columnwidth]{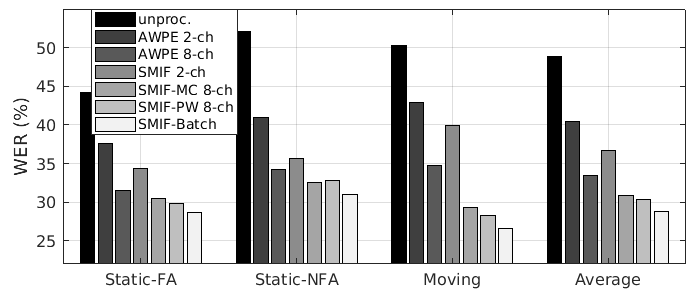}} 
\caption{\small  Dereverberation performance, i.e. SRMR, PESQ, STOI metrics and WER (from top to bottom), for the Dynamic dataset. The WER of close-talk signals for the three subsets are 22.1\%, 24.2\% and 14.4\%, respectively. } 
\label{fig:dynres}
\vspace{-0.0cm}
\end{figure}

Fig.~\ref{fig:dynres} presents the dereverberation results for the three subsets in the Dynamic dataset. For the unprocessed data, all the performance measures are bad due to the intense reverberation. The Static-NFA set has the lowest SRMR and PESQ metrics. When speakers do not face the microphones, the direct-path speech signal received by microphones becomes smaller relative to the reverberation and ambient noise, in other words the microphone signals are more reverberated and noisy. The Moving case has the lowest STOI metrics. The WER clearly increases from the Static-FA set to the Static-NFA and Moving sets.   

For all conditions and performance metrics, the proposed 8-ch SMIF-MC and SMIF-PW methods perform similarly, thence we will not  distinguish them in the following. 
For both AWPE and the proposed method, the SRMR performance slightly degrades from the Static-FA set to the Static-NFA set, and further noticeably degrade for the Moving set. AWPE achieves larger PESQ metrics than the proposed method for the static cases, but has a large performance degradation for the Moving set. By contrast, the proposed method achieves even larger PESQ metrics for the Moving set. In terms of STOI, the two methods perform similarly for the static cases, and the proposed method outperforms AWPE for the Moving set. 
As for ASR, the proposed method outperforms AWPE, especially for the Moving set. 
Overall, the performance measures show the comparable dereverberation capability of AWPE and the proposed method  for the static speaker cases, and show the   superiority of the proposed method for the moving speaker case.
The Dynamic dataset is more challenging than the REVERB dataset in terms of  adaptive (inverse) filter estimation mainly due to its lower DRR and $C_{50}$. In addition, the moving speaker case suffers a larger filter estimation error compared to the static speaker case, due to the imperfect tracking ability.  Compared to the complex-valued inverse filtering in AWPE, the proposed STFT magnitude inverse filtering is less sensitive to additive noise, filter perturbations and other unexpected distortions \cite{kinoshita2016}.

The batch mode counterpart of the proposed method, referred to as SMIF-Batch in Fig. 2, uses eight microphones and the pairwise scheme of magnitude inverse filtering.  \addnote[batchreason]{1}{The speech enhancement and ASR performance measures of the batch method are not consistent. Compared to the online method,  on the one hand, the batch method achieves worse speech enhancement metrics, even for the static speaker cases. On the other hand, it performs slightly better for ASR, even for the moving speaker case. The reason for this inconsistency is not very clear. The present work uses the critically sampled CTF convolution and the magnitude CTF convolution, which are rough approximations. As a result, for the static speaker case, the CTF and inverse filter that optimize the approximations are actually time-varying, and thus the online method could sometimes outperform the batch method.} 

\begin{figure}[t]
\centering
{\includegraphics[width=0.95\columnwidth]{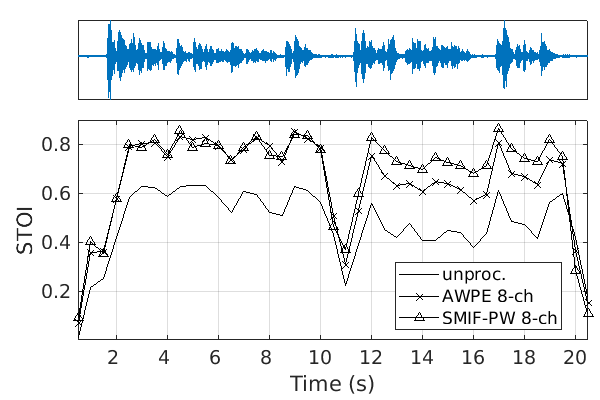}}
\caption{\small The short-time STOI metrics computed with a 1-s sliding window and 0.5 s sliding step. One speaker was standing at one point within 0-11 s, and started walking to another point from 11 s. } 
\label{fig:example}
\vspace{-0.0cm}
\end{figure}

\begin{figure*}[t]
\centering
\subfloat[]{\includegraphics[width=0.48\textwidth]{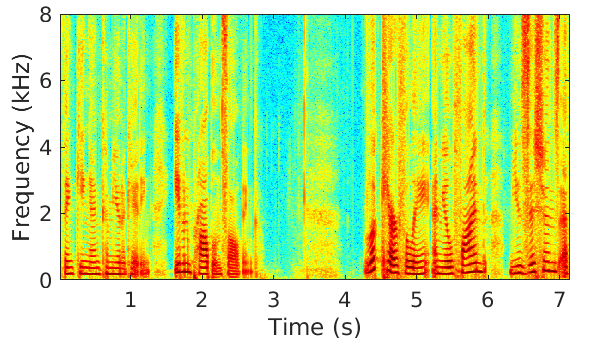}} 
\subfloat[]{\includegraphics[width=0.48\textwidth]{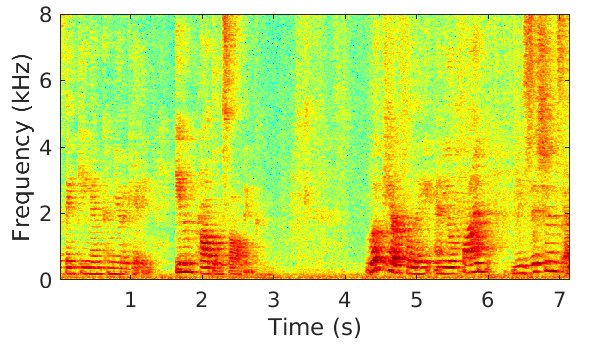}} \\
\subfloat[]{\includegraphics[width=0.48\textwidth]{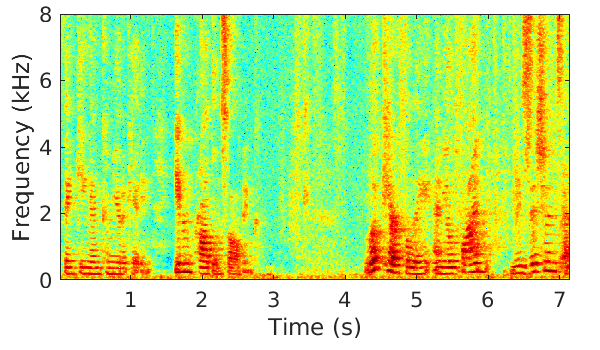}}  
\subfloat[]{\includegraphics[width=0.48\textwidth]{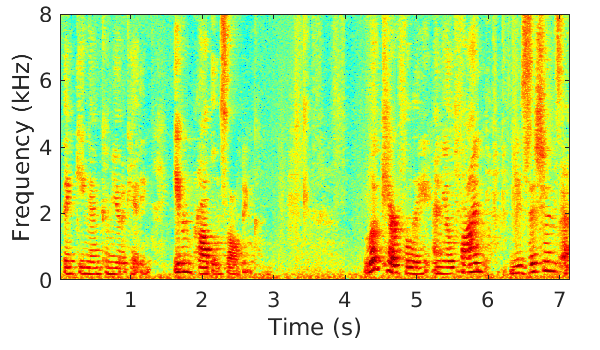}} 
\caption{\small Example of spectrogram for a signal from the Dynamic dataset. (a) close-talk clean signal, (b) microphone signal, (c)  enhanced signal by 8-ch AWPE and (d) the proposed 8-ch SMIF-PW method. The speaker was static with in 0-4 s, and started walking from one point to another from 4 s. } 
\label{fig:spectrogram}
\vspace{-0.2cm}
\end{figure*}

Fig.~\ref{fig:example} shows the STOI metrics computed with a 1-s sliding window for one audio recording. This result is consistent with Fig.~\ref{fig:dynres} depicting  that the two methods have comparable  STOI metrics when the speaker is static before 11 s, and the proposed method achieves higher STOI metrics when the speaker is moving after 11 s. When the speaker starts speaking after a silent period, the two methods adapt from background noise to speech, and quickly converge. It is observed from Fig.~\ref{fig:example} that the two methods have a similar convergence speed, i.e. less than 1 s.  Fig.~\ref{fig:spectrogram} depicts the spectrograms of the middle part  (around the point where the speaker starts moving) of the recording in Fig.~\ref{fig:example}. It can be seen that reverberation is largely removed by both methods. However, the difference between the two methods and the difference between the static and moving cases cannot be clearly observed from the spectrograms. 
Informal listening tests show that, the proposed method is not perceived to have more residual reverberation for the moving speaker case compared to the static speaker case. 
Audio examples for all experiments presented in this paper are available in our website.\footnote{https://team.inria.fr/perception/research/ctf-dereverberation}

\subsection{Computational Complexity Analysis}
Both the proposed method and AWPE are frame-wise online methods. We analyze their computational complexity for one frame.
The proposed method consists of CTF identification and magnitude inverse filtering. The computation of CTF identification is mainly composed of Algorithm~\ref{alg:sm}, which executes (\ref{eq:sm}) $I-1$ times. The computation of (\ref{eq:sm}) includes three matrix-vector multiplications. The matrix/vector size is $I\tilde{Q}$. We remind that $I=2 \ \text{or} \ 8$ and $\tilde{Q}=4$ are the number of channels  and  the length of the critically sampled CTF, respectively.  CTF identification is performed for each of the $N/2+1$ positive-valued frequency bins.  Overall, the computational complexity of CTF identification is approximately $\mathcal{O}(NI^3\tilde{Q}^2)$. The computation of inverse filtering is mainly composed of the gradient calculation (\ref{eq:gra}), which includes two matrix-vector multiplications. However, each of these  multiplications actually represents $I$ one-dimensional convolutions. In practice, we implement the convolution using an FFT (fast Fourier transform) with $N_{\text{fft}}=2\tilde{Q}+\tilde{O}-2$ points, where $\tilde{O}=4$ is the length of the inverse filter. Overall, the computational complexity of multichannel inverse filtering is approximately $\mathcal{O}(NIN_{\text{fft}}\text{log}(N_{\text{fft}}))$. For the pairwise processing scheme, the two-channel inverse filtering is executed $I(I-1)/2$ times, thence the computational complexity is $\mathcal{O}(NI^2N_{\text{fft}}\text{log}(N_{\text{fft}}))$. 

 \setlength{\tabcolsep}{5.0pt}
\begin{table}[h]
\centering
\caption{\small Real-time factor for AWPE and each step of the proposed method.}
\label{tab:rf}
\begin{tabular}{l l | c   c   }   
	Method  &  &  2-ch   &  8-ch     \vspace{1mm} \\  \hline 
\multicolumn{2}{l|}{AWPE}        & 0.54  & 2.45  \vspace{1mm}  \\ 
SMIF  & CTF identification       & 0.11  & 2.73 	\\
(ours)     & Inverse filtering (SMIF-MC)  & 0.09 & 0.52 \\
     & Inverse filtering (SMIF-PW)  & 0.09 & 1.35  \\
     & Overall (SMIF-MC) & 0.20 & 3.25                  \\
     & Overall (SMIF-PW) & 0.20 & 4.08
\end{tabular}
\end{table} 

Similar to the proposed CTF identification method, the computation of RLS-based AWPE is also composed of matrix-vector multiplications. The matrix/vector size is $IQ_{\text{wpe}}$, where $Q_{\text{wpe}}$ denotes the length of the prediction filter, i.e.  16 and 8 for the 2-ch and 8-ch cases, respectively. The computational complexity of AWPE is $\mathcal{O}(N_{\text{wpe}}I^2Q_{\text{wpe}}^2)$, where $N_{\text{wpe}}$ denotes the STFT frame length for AWPE, i.e. 512 in this experiment. 
 
The computation time is measured with the real-time factor (RF), which is the processing time of a method divided by the length of the processed signal. Both AWPE and the proposed method are implemented in MATLAB. RF for WPE and each step of the proposed method are shown in Table~\ref{tab:rf}. For the 2-ch case,  all processes have an RF smaller than 1,  and thus can be run in real-time. The proposed method is less time-consuming than AWPE, since the critically sampled CTF and inverse filter of the proposed method are shorter than the predicition filter of AWPE, i.e. 4 versus 16. For the 8-ch case, AWPE is faster than the proposed method. As analyzed above, the computational complexity of the proposed CTF identification is cubic of the number of channels, while the one of AWPE is square of the number of channels.

\section{Conclusions}\label{sec:conclusion}
In this paper, a blind multichannel online dereverberation method has been proposed. The batch algorithm for multichannel CTF identification  proposed in our previous work \cite{li2018taslp2} was extended to an online method based
on the RLS criterion. Then, a gradient descent-based adaptive magnitude MINT was proposed to estimate the inverse filters of the identified CTF magnitude. Finally, an estimate of the STFT magnitude of the source signal can be obtained by applying the inverse filtering onto the STFT magnitude of the microphone signals. 
Experiments were conducted in terms of both speech quality and intelligibility. Compared to the AWPE method, the proposed method achieves comparable ASR performance on the REVERB challenge dataset. Experiments with the Dynamic dataset show that the proposed method performs better than AWPE for the moving speaker case due to the robustness of the STFT magnitude-based scheme. \addnote[noise]{1}{Even though the proposed method does not account for noise reduction at all, the dereverberation experiments were performed on data including additive noise. The experimental results indicate that the dereverberation capability of the proposed method is not significantly deteriorated by the additive noise. However, the noise in the dereverberated signal still has some influence on both human listening and ASR metrics. A noise reduction method that fits well the proposed dereverberation method will be investigated in the future.} 

\bibliographystyle{ieeetr}
\balance

\end{document}